\documentclass[fp,twocolumn]{jpsj3}
\usepackage{txfonts}
\usepackage{autobreak}

\usepackage{graphicx}
\usepackage{bm}
\usepackage{color}
\usepackage{mathrsfs}

\title{Two-band Model with High Thermoelectric Power Factor and Its Application to FeSe Thin Film} 

\author{Manaho Matsubara$^1$\thanks{mmatsubara@rs.tus.ac.jp}, Takahiro Yamamoto$^{1,2}$ and Hidetoshi Fukuyama$^{2}$}
\inst{$^1$Department of Physics, Faculty of Science, Tokyo University of Science, Shinjuku, Tokyo 162-8601, Japan \\
$^2$ Research Institute for Science and Technology, Tokyo University of Science, Shinjuku, Tokyo 162-8601, Japan
} 

\abst{
We propose a simple theoretical model referred to as the {\it two-band model} to realize both a large Seebeck coefficient and high electrical conductivity, resulting in a high thermoelectric (TE) power factor ($PF$).
Using the Kubo--Luttinger linear response theory, we apply this model to the TE response of an FeSe thin film reported by Shimizu {\it et al.} to show a high $PF$ with strong temperature dependence.
The effects of superconducting fluctuations and excitonic correlations are found to not be critical.
}

\begin{document}
\maketitle

\section{Introduction}
Because of recent societal needs for energy harvesting technology, the development of high-performance thermoelectric (TE) materials has been a topic of intense interest.
The performance of a TE material is characterized by its power factor $PF=L_{11}S^2$ and the dimensionless figure of merit $ZT=(PF/\lambda)T$; the former represents the maximum output, and the latter represents the maximum efficiency.
Because the $PF$ and $ZT$ are determined by three intercorrelated physical quantities ({\it i.e.}, the electrical conductivity $L_{11}$, the Seebeck coefficient $S$, and the thermal conductivity $\lambda$), enhancing the performance of TE materials is difficult.
Nevertheless, extensive effort has been devoted to developing materials with enhanced $PF$ and $ZT$ values, and several high-performance TE materials with $ZT>2$ have recently been reported~\cite{rf:mori}.
In addition, new TE materials with high $PF$ values ({\it e.g.}, organic materials~\cite{rf:organic} and carbon nanotubes~\cite{rf:cnt}) have also recently been developed.
To attain TE materials with even higher performance, we propose a theoretical model with a high $PF$, which is made possible by modulating electronic states.

In principle, the $PF$ can be large even for a moderate $S$ if the $L_{11}$ is sufficiently high in systems with many valleys with degeneracy, $n_{\rm v}$, because $L_{11}$ is proportional to $n_{\rm v}$, whereas $S$ is independent of $n_{\rm v}$.
This is not easy to realize, however, and then efforts to find systems with a large (absolute value of) $S$ have been pursued~\cite{{rf:largeS1},{rf:largeS2},{rf:largeS3}}.
Because $S$ is dependent on the details of a material's electronic state, collaborative theoretical--experimental studies through searches for new materials, experimental characterization of their electronic properties, and detailed theoretical analysis are strongly desired.
(This situation is similar to efforts to realize superconductors with a high critical temperature, $T_{\rm c}$.)
In pursuing this cycle, we expect that focusing on known cases with particular features and then analyzing them in detail to extract essential properties of the electronic states underlying the experimental facts will guide further efforts to develop new materials.
In the present paper, we propose a simple theoretical model, the {\it two-band model}, to realize both a large $S$ and high $L_{11}$, leading to a high $PF$.
We then apply it to recent experimental data for an FeSe thin film with a high $PF$, as reported by Shimizu {\it et al.}~\cite{rf:shimizu}.

Shimizu {\it et al.} have reported that an FeSe thin film shows an unprecedented large $PF$ value in the presence of a vertical electric field $\mathcal E_{\perp}$ over a wide temperature ($T$) range ($50~{\rm K}<T<300~{\rm K}$)~\cite{rf:shimizu}.
The FeSe thin film exhibits strongly metallic character.
Its $L_{11}$ increases with decreasing $T$ below room $T$, with the $L_{11}\sim 4\times 10^{6}$~S/m at $T\sim 100$~K and the film eventually becoming superconducting below $T_{\rm c} \sim 50$~K. By contrast, $S$ shows weak $T$ dependences for $50~{\rm K}<T<200~{\rm K}$, with a large value of 350~$\mu$V/K, leading to a $PF$ of $\sim 500$~mW/(m$\cdot$K$^2$) at $T \sim 100$~K~\cite{rf:shimizu}.

Bulk FeSe consists of two-dimensional FeSe layers and is known to be a semimetal~\cite{{rf:shimizu},{rf:fese_band1}}.
The electron and hole energy bands at the Fermi energy consist mainly of the $d$ orbitals of Fe, and both bands are doubly degenerate.
Similar to bulk FeSe, monolayer FeSe is also a semimetal~\cite{{rf:fese_band2}}.
Such a semimetal does not usually show a large $S$ similar to that of a semiconductor.
Thus, we speculate that the large $S$ of FeSe thin films originates from the effects of the $\mathcal E_{\perp}$ applied to thin films.
The $\mathcal E_{\perp}$ would not only open the bandgap of FeSe but also split its degenerate energy bands (Fig.~1).
In fact, Shimizu {\it et al.} have reported an experimental investigation of the field-induced bandgap of FeSe thin films~\cite{rf:shimizu}. 
\begin{figure}[t]
  \begin{center}
  \includegraphics[keepaspectratio=true,width=90mm]{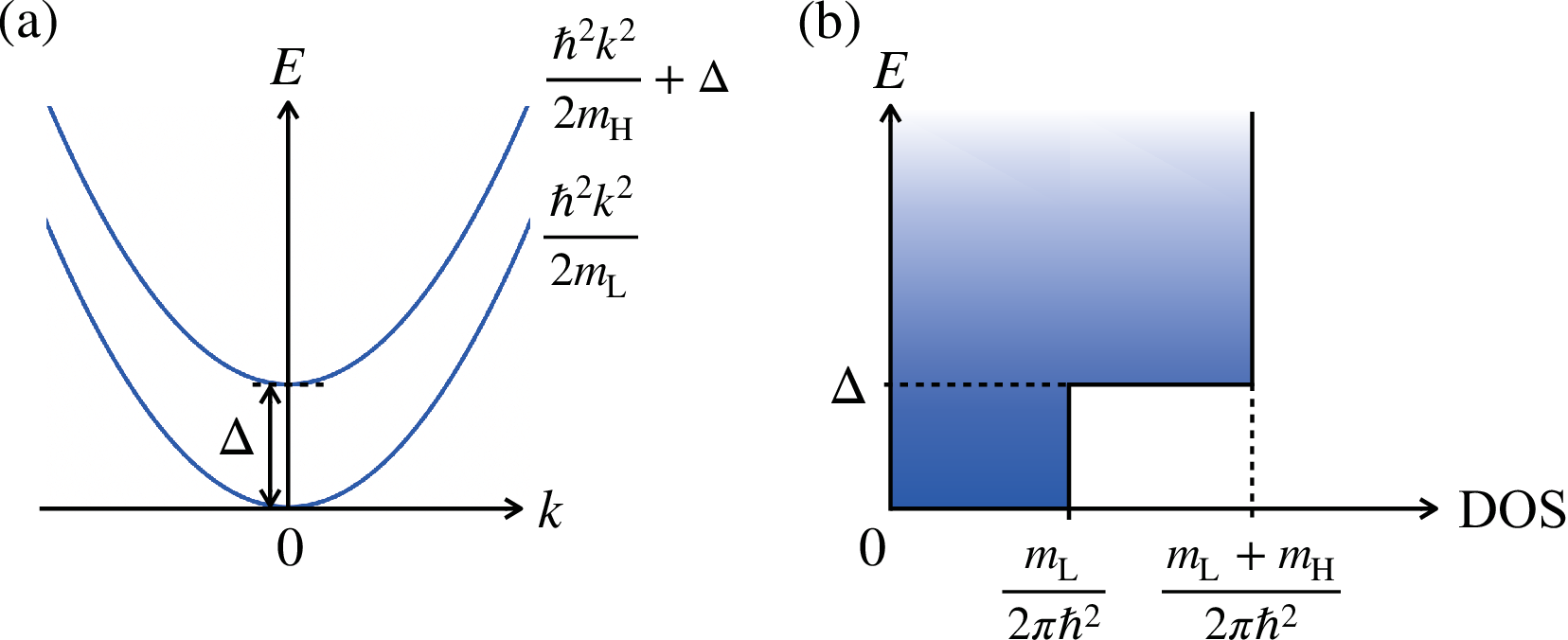}
  \end{center}
\caption{(Color online) (a) Higher and lower energy bands with energy spacing $\Delta$.
(b) Density of states (DOS) for the two-band model in two dimensions.
Here, $m_{\rm H/L}$ is the effective mass of the higher/lower-band electrons.}
\label{fig:01}
\end{figure}

In the present paper, we show that the two-band model achieves a high $PF$. 
We then analyze the experimental data for an FeSe thin film on the basis of the proposed model.
The TE response functions are calculated according to Kubo--Luttinger (KL) theory combined with the Green's function method, which is a fully quantum theory for TE transport in materials~\cite{{rf:kubo},{rf:Luttinger}}.

\section{Model and Theory~\label{sec:2}}
In this section, we explain the KL theory (\S~2.1) and introduce the two-band model (\S~2.2).

\subsection{Kubo--Luttinger theory of TE response~\label{sec:2.1}}
We explain the KL theory combined with the Green's function method, which enables the TE response to be analyzed fully quantum mechanically~\cite{{rf:kubo},{rf:Luttinger}}.

The current density $J$ in the presence of both an electric field $\mathcal{E}$ and a temperature gradient $dT/dx$ in the $x$-direction in the material is described as 
\begin{eqnarray}
J=L_{11}\mathcal{E}-\frac{L_{12}}{T}\frac{dT}{dx}
\label{eq:J}
\end{eqnarray}
within the linear response to $\mathcal{E}$ and $dT/dx$.
Here, $L_{11}$ and $L_{12}$ are the electrical conductivity and the thermoelectrical conductivity, respectively.
Using $L_{11}$ and $L_{12}$, we can express $S$ as
\begin{eqnarray}
S=\frac{1}{T}\frac{L_{12}}{L_{11}}
\label{eq:S}
\end{eqnarray}
and can describe the $PF$ by
\begin{eqnarray}
PF= L_{11} S^2=\frac{1}{T^2}\frac{L_{12}^2}{L_{11}}.
\label{eq:PF}
\end{eqnarray}
According to the KL theory, $L_{11}$ and $L_{12}$ are given by 
\begin{align}
L_{11}&\propto \int_{0}^\beta \!\!\!d\tau\left\langle T_\tau \{J_{\rm el}(\tau)J_{\rm el}(0)\}\right\rangle e^{i\omega_\lambda\tau},
\label{eq:L11_KL}\\
L_{12}&\propto \int_{0}^\beta \!\!\!d\tau\left\langle T_\tau \{J_{\rm el}(\tau)J_{\rm th}(0)\}\right\rangle e^{i\omega_\lambda\tau},
\label{eq:L12_KL}
\end{align}
where $J_{\rm el}$ and $J_{\rm th}$ are electrical and thermal current operators, respectively.
On the basis of the KL theory together with the method introduced by Jonson and Mahan~\cite{rf:JM}, Ogata and Fukuyama~\cite{rf:MO_HF} clarified that $L_{11}$ and $L_{12}$ can be expressed as the Sommerfeld--Bethe relation~\cite{rf:sommerfeld},
\begin{align}
L_{11}&=\int_{-\infty}^\infty \!\!\!dE\left(-\frac{\partial f(E-\mu)}{\partial E}\right)\alpha(E),
\label{eq:L11}\\
L_{12}&=-\frac{1}{e}\int_{-\infty}^\infty \!\!\!dE\left(-\frac{\partial f(E-\mu)}{\partial E}\right)(E-\mu)\alpha(E),
\label{eq:L12}
\end{align}
even for strongly disordered systems (with particular types of both mutual electron interactions and electron--phonon interactions)~\cite{{rf:TY_HF},{rf:TY_HF2},{rf:matsubara},{rf:horii},{rf:YOF}}.
Here, $\alpha(E)$ is the spectral conductivity, $e$ is the elementary charge, $\mu$ is the chemical potential, and $f(E-\mu)=1/(\exp((E-\mu)/k_{\rm B}T)+1)$ is the Fermi--Dirac distribution function.

\subsection{Two-band model for an FeSe thin film~\label{sec:2.2}}
\subsubsection{Effective Hamiltonian}
We introduce a two-band model in two dimensions that is applicable as an effective Hamiltonian for an FeSe thin film in $\mathcal E_{\perp}$.
The Hamiltonian of the band model is given by
\begin{eqnarray}
H=g_{\rm s}g_{\rm o}\sum_{\bm k}\left\{
\epsilon_{\rm L}(\bm k)c^\dagger_{{\rm L}{\bm k}}c_{{\rm L}{\bm k}}+\epsilon_{\rm H}(\bm k)c^\dagger_{{\rm H}{\bm k}}c_{{\rm H}{\bm k}}
\right\},
\label{eq:h}
\end{eqnarray}
where $\epsilon_{\rm L}=\frac{\hbar^2k^2}{2m_L}$ and $\epsilon_{\rm H}=\frac{\hbar^2k^2}{2m_H}+\Delta$ are the energy dispersions for the lower and higher conduction bands, respectively, with an energy spacing $\Delta$ (Fig.~\ref{fig:01}(a)). Here, $c^\dagger_{{\rm L}{\bm k}}$ and $c^\dagger_{{\rm H}{\bm k}}$ ($c_{{\rm L}{\bm k}}$ and $c_{{\rm H}{\bm k}}$) are the creation (annihilation) operators for the lower and higher conduction-band electrons with the two-dimensional wavenumber ${\bm k}=(k_x, k_y)$.
$m_{\rm L}$ and $m_{\rm H}$ are the effective masses of the lower and higher conduction-band electrons, respectively.
$g_{\rm s}$ and $g_{\rm o}$ are the spin and orbital degrees of freedom, respectively.
For an FeSe thin film, they are given by $g_{\rm s}=2$ and $g_{\rm o}=1$.
The origin of energy is taken to be the bottom of the lower conduction band. 

\begin{figure}[t]
  \begin{center}
  \includegraphics[keepaspectratio=true,width=80mm]{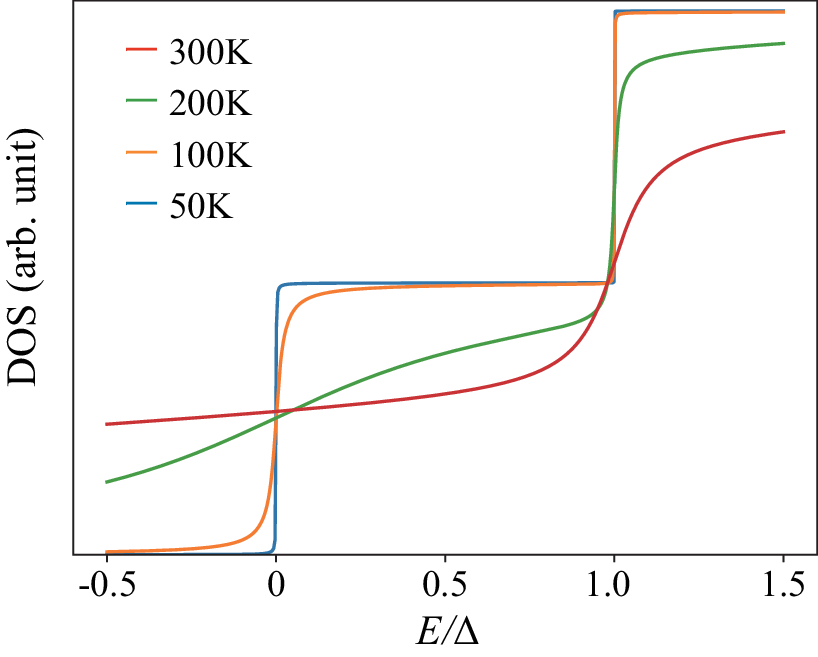}
  \end{center}
\caption{(Color online) Effective density of states in the presence of scattering for the two-band model for several temperatures: $T=50$, 100, 200, and 300 K.
Here, $\Delta$ is the energy spacing between the higher- and lower-band bottoms.}
\label{fig:rho}
\end{figure}

\subsubsection{Green's function~\label{sec:2.2}}
We introduce a retarded Green's function as
\begin{eqnarray}
G_{\rm L/H}^{\rm R}(E,{\bm k})=\frac{1}{E-\epsilon_{\rm L/H}-\Sigma_{\rm L/H}^{\rm R}},
\label{eq:green}
\end{eqnarray} 
where $\Sigma^{\rm R}_{\rm L/H}$ is the retarded self-energy due to several scattering events for the lower/higher conduction-band electrons.
In the present study, we adopt the constant-$\tau$ approximation with respect to energy $E$ and wavenumber ${\bm k}$ as
\begin{eqnarray}
\Sigma_{\rm L/H}^{\rm R}=-i\frac{\hbar}{2\tau_{\rm L/H}}
\label{eq:sigma}
\end{eqnarray}
and consider electron--phonon interaction and impurity scattering as scattering processes in the system.
${\hbar}/{\tau_{\rm L/H}}$ is assumed to be
\begin{eqnarray}
\frac{\hbar}{\tau_{\rm L/H}(T)}&=A_{\rm L/H}T^5+B_{\rm L/H},
\label{eq:tau}
\end{eqnarray}
where the first term with parameter $A_{\rm L/H}$ is the contribution from the electron--phonon scattering; the Bloch-Gr{\" u}neisen law (the $T^5$-law) is assumed,
and the second term with the parameter $B_{\rm L/H}$ is the contribution from the impurity scattering, which is independent of $T$.

\subsubsection{Electronic state: Density of states~\label{sec:2.2}}

The density of states (DOS) $\rho(E)$ per unit area in the presence of scattering for the two-band model is expressed as 
\begin{eqnarray}
\rho(E)=-\frac{g_{\rm s}g_{\rm o}}{\pi\Omega}\sum_{\bm k}\left({\rm Im} G_{\rm L}^{\rm R}(E,{\bm k}) + {\rm Im} G_{\rm H}^{\rm R}(E,{\bm k})\right)
\label{eq:self-energy01}
\end{eqnarray}
in terms of the retarded Green's function.
Here, $\Omega$ is the system area.
Substituting $G_{\rm L/H}^{\rm R}(E,{\bm k})$ in Eqs.~(\ref{eq:green}) and (\ref{eq:sigma}) into Eq.~(\ref{eq:self-energy01}), we can analytically calculate $\rho(E)$ as
\begin{eqnarray}
\rho(E)=\frac{g_{\rm s}g_{\rm o}}{2\pi^2\hbar^2}
\left\{
m_{\rm L}(\pi-\theta_{\rm L})+m_{\rm H}(\pi-\theta_{\rm H})
\right\},
\label{eq:self-energy}
\end{eqnarray}
where $\theta_{\rm L/H}$ are defined in terms of the relaxation times $\tau_{\rm L/H}$,
\begin{eqnarray}
\theta_{\rm L}&=&\arctan\frac{\hbar}{2\tau_{\rm L}E}\ \ \ \ \ \ \ \ \ \ (0\leq\theta_{\rm L}<\pi),\\
\label{eq:theta1}
\theta_{\rm H}&=&\arctan\frac{\hbar}{2\tau_{\rm H}\left(E-\Delta\right)}\ \ \ (0\leq\theta_{\rm H}<\pi).
\label{eq:theta2}
\end{eqnarray}
Here, the DOS in Eq.~(\ref{eq:self-energy}) is effective one which depends on $T$ through $\tau_{\rm L/H}$ in Eq.~(\ref{eq:tau}). 
Figure~\ref{fig:rho} shows $\rho(E)$ of the two-band model at several temperatures for the choice of $m_{\rm L}=m_{\rm H}=2.7m_0$ ($m_0$: electron mass in vacuum), $\Delta = 29$~meV, $A_{\rm L}=1.0\times 10^{-13}$~eV/K$^5$, $A_{\rm H}=2\times 10^{-15}$~eV/K$^5$, and $B_{\rm L}=B_{\rm H}=0$.
Notably, these parameters are determined to reproduce the experimental data for an FeSe thin film, as will be shown later (\S~3.1).
The solid blue, orange, green, and red curves correspond to $\rho(E)$ at $T=50$, $100$, $200$, and 300~K, respectively.
The blue curve ($T=50$~K) has a stepwise structure with a sharp step reflecting the two-dimensionality of the present system.
As $T$ increases, the stepwise DOS become smoother because $\tau_{\rm L/H}$ become short and eventually $\theta_{\rm L/H}$ in Eqs.~(14) and (15) become large.
In Fig.~\ref{fig:rho}, compared with the higher-band bottom, the lower-band bottom has a greater scattering effect on the DOS.

\begin{figure}[t]
  \begin{center}
  \includegraphics[keepaspectratio=true,width=80mm]{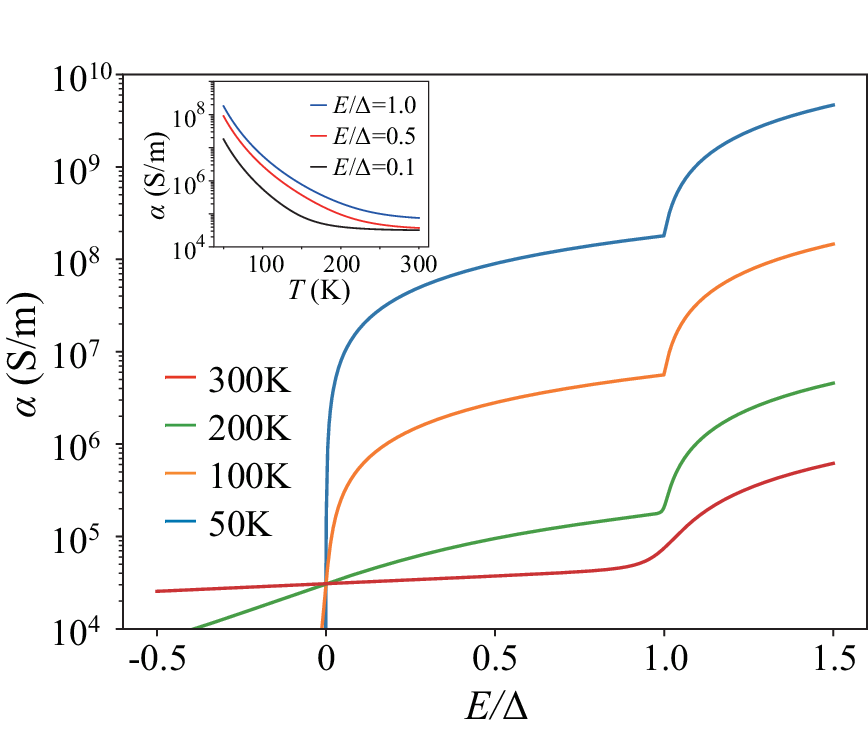}
  \end{center}
\caption{(Color online) Spectral conductivity $\alpha$ for several temperatures: $T=50$, $100$, $200$, and $300$~K.
The inset denotes the $T$ dependences of $\alpha$ for several fixed energies: $E/\Delta=0.1$, $0.5$, and $1.0$ with $\Delta=29$~meV.}
\label{fig:alpha}
\end{figure}

The spectral conductivity $\alpha(E)$ is described in terms of the Green's functions as
\begin{eqnarray}
\alpha(E)=g_{\rm s}g_{\rm o}\frac{e^2 \hbar }{\pi V}\sum_{\bm k}
\left(v_{\rm L}^2\left[{\rm Im}G_{\rm L}^{\rm R}(E,{\bm k})\right]^2
+v_{\rm H}^2\left[{\rm Im}G_{\rm H}^{\rm R}(E,{\bm k})\right]^2\right),
\end{eqnarray}
where $v_{\rm L/H}=\hbar k_x/m_{\rm L/H}$ is the group velocity in the $x$-direction and $V$ is the system volume.
The summation with respect to ${\bm k}$ can be analytically performed; $\alpha(E)$ is eventually given by
\begin{eqnarray}
\alpha(E)
&=&g_{\rm s}g_{\rm o}\frac{e^2}{4\pi^2\hbar d}
\left[\left\{\frac{2\tau_{\rm L}}{\hbar}E(\pi-\theta_{\rm L})+1\right\}\right.\nonumber\\
& &+\left.\left\{\frac{2\tau_{\rm H}}{\hbar}(E-\Delta)(\pi-\theta_{\rm H})+1\right\}
\right],
\label{eq:alpha}
\end{eqnarray}
where $d=4~\AA$ is the distance between the atomic layers in the system.
Here, we note that $\alpha(E)$ is independent of the effective masses $m_{\rm H}$ and $m_{\rm L}$, which is a characteristic feature of two-dimensional electron systems.
Figure~\ref{fig:alpha} shows $\alpha(E)$ for several temperatures: $T=50$, $100$, $200$, and $300$~K. $\alpha(E)$ increases stepwise with increasing $E$, reflecting the DOS.
The inset in Fig.~\ref{fig:alpha} denotes the $T$ dependences of $\alpha(E)$ for several fixed energies: $E/\Delta=0.1$, $0.5$, and $1.0$.
The figure shows that $\alpha(E)$ with a fixed $E$ increases almost exponentially as $T$ decreases.

\section{Simulation Results~\label{sec:3}}

\begin{figure}[t]
  \begin{center}
  \includegraphics[keepaspectratio=true,width=80mm]{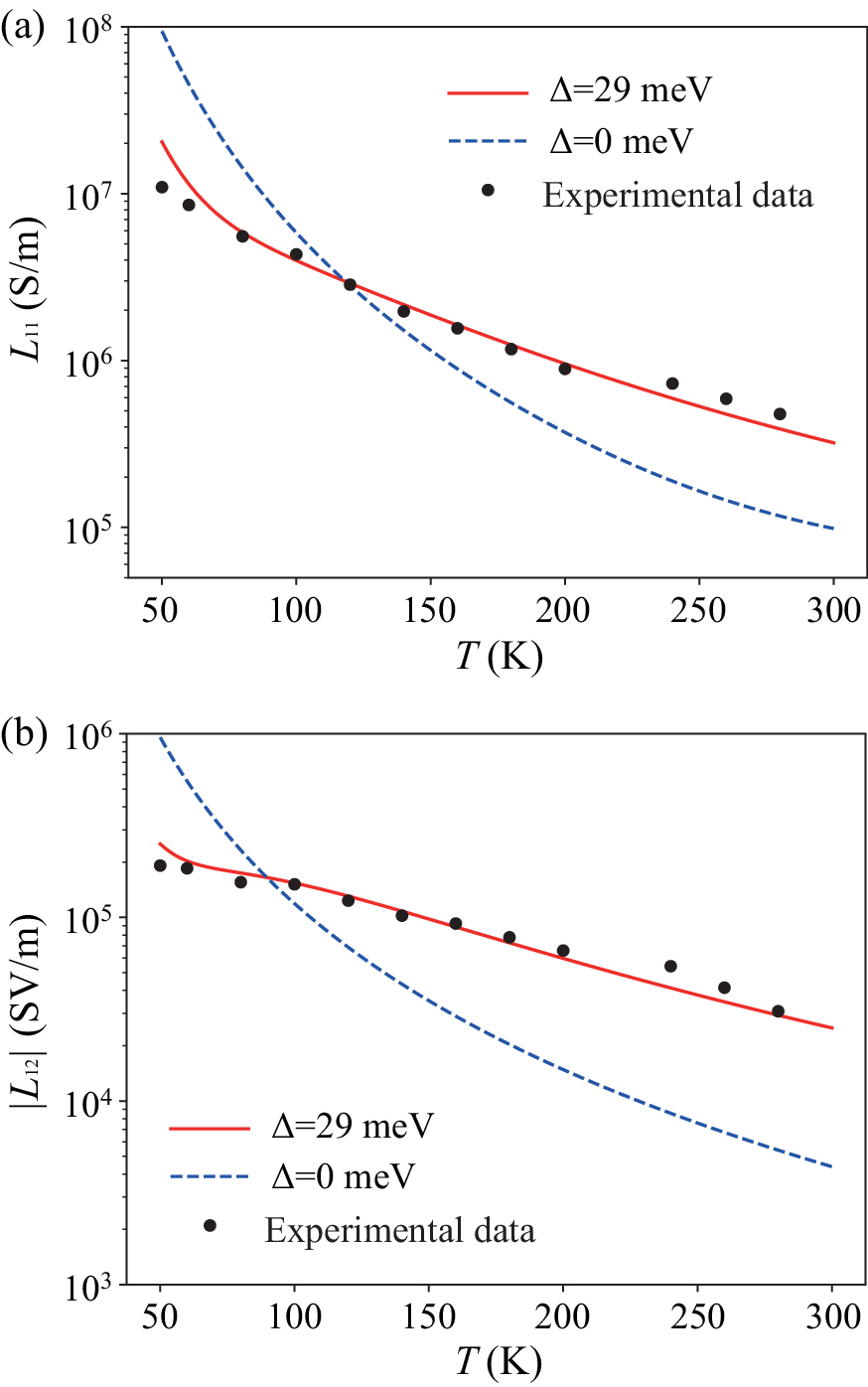}
  \end{center}
\caption{(Color online) $T$ dependences of (a) $L_{11}$ and (b) $L_{12}$ for an FeSe thin film.
The solid circles are experimental data in Ref.~\citen{rf:shimizu}.
The solid curve represents the theoretical curve calculated on the basis of the two-band model with the energy spacing $\Delta=29$~meV.
The dashed curve denotes the theoretical curve calculated on the basis of the degenerate two-band model with $\Delta=0$ eV.}
\label{fig:l11}
\end{figure}

\subsection{Temperature dependence of $L_{11}$ and $L_{12}$}
Figures~\ref{fig:l11}(a) and \ref{fig:l11}(b) represent the $T$ dependence of $L_{11}$ and $L_{12}$ of the two-band model from 50~K to 300~K, along with the experimental data for an FeSe thin film, where $T_{\rm c}=50$~K is the superconducting transition temperature of the FeSe thin film.
The solid circles are experimental data in the case of finite gate voltage $V_{\rm G}=3.95$~V reported by Shimizu {\it et al.}~\cite{rf:shimizu}.
$L_{11}$ and $|L_{12}|$ decay strongly with increasing $T$ because of the electron--phonon scattering.
The solid curves in Figs.~\ref{fig:l11}(a) and \ref{fig:l11}(b) are theoretical curves calculated on the basis of the two-band model with a finite split.
The parameters in the model are determined by fitting the experimental data: $\Delta=29$~meV, $\mu=0.1$~meV, $A_{\rm L}=1.0\times 10^{-13}$~eV/K$^5$, $A_{\rm H}=2\times 10^{-15}$~eV/K$^5$, and $B_{\rm L}=B_{\rm H}=0$~eV.
The results indicate that the present system is a good metal and that the electron--phonon scattering is dominant in comparison with the impurity scattering within the present $T$ region. By contrast, $A_{\rm H}<A_{\rm L}$ indicates that the lower-band electrons are more affected by electron--phonon scattering than the higher-band electrons.
For reference, $A=5\times 10^{-13}$~eV/K$^5$ and $B=3\times 10^{-5}$~eV for copper~\cite{rf:copper}.
If the two bands are assumed to be degenerate, then the best-fit parameters are $\mu=0.1$~meV and $A_{\rm L}=A_{\rm H}=0.4\times 10^{-13}$~eV/K$^5$ for $B_{\rm L}=B_{\rm H}=0$~eV, as in the two-band model.
The results show that the experimental data cannot be reproduced by the degenerate two-band model with $\Delta=0$~eV.
As we will explain in the next section, to realize both a large $S$ and high $L_{11}$, it is critical that the chemical potential ($\mu=0.1$~meV ) be located between the higher and lower conduction bands and that $A_{\rm H}$ be smaller than $A_{\rm L}$.
By contrast, the experimental data in Ref.~\citen{rf:shimizu} could not be reproduced under the condition of $A_{\rm H}=A_{\rm L}$ for any $\Delta$.
\begin{figure}[t]
  \begin{center}
  \includegraphics[keepaspectratio=true,width=80mm]{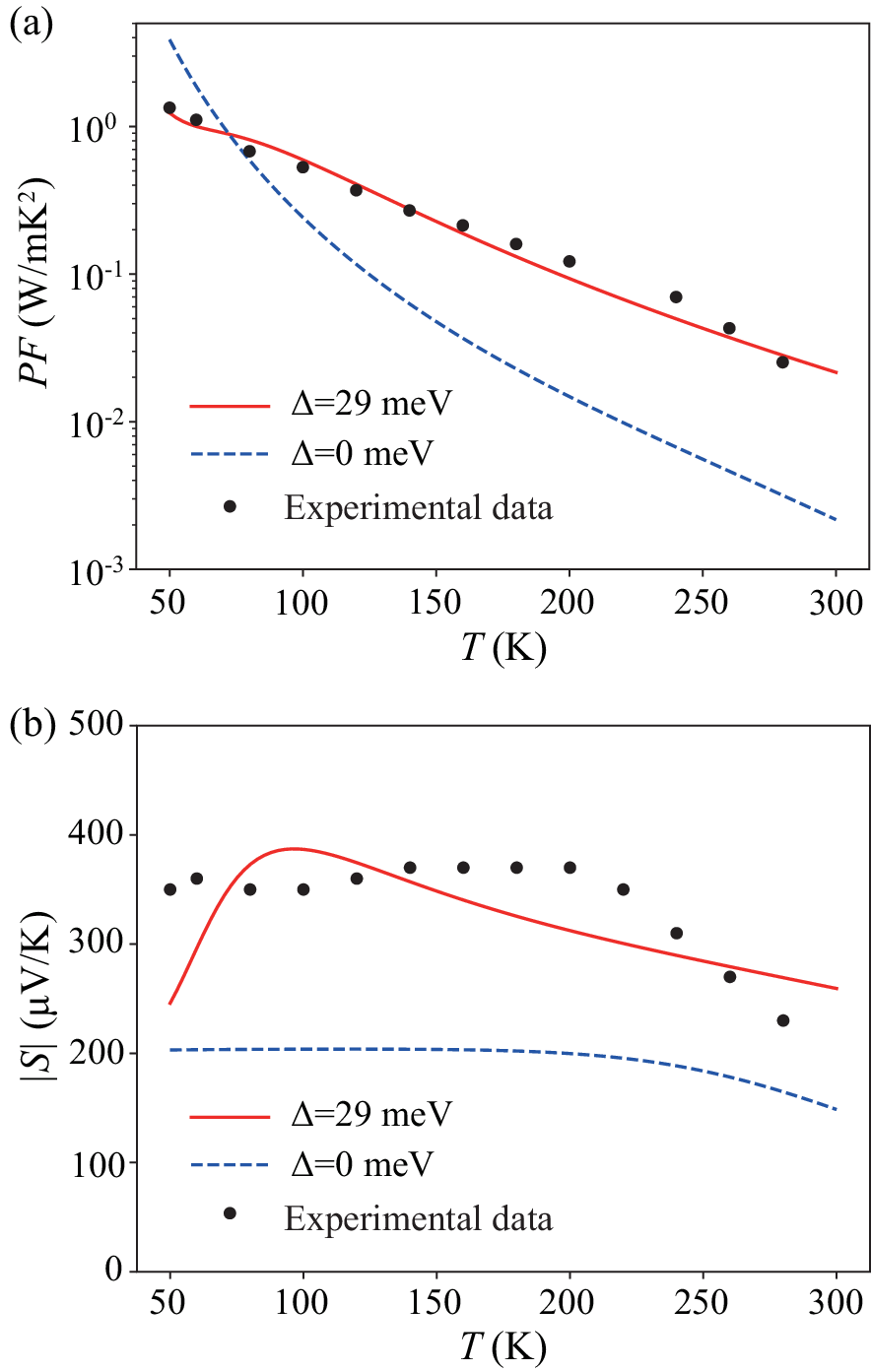}
  \end{center}
\caption{(Color online) $T$ dependences of the (a) $PF$ and (b) $S$ in the two-band mode, along with the experimental data for an FeSe thin film.
The solid circles are experimental data reported in Ref.~\citen{rf:shimizu}.
The solid curve represents the theoretical curve calculated on the basis of the two-band model with the energy spacing $\Delta=29$~meV.
The dashed curve denotes the theoretical curve calculated on the basis of the degenerate two-band model with $\Delta=0$ eV.}
\label{fig:pf}
\end{figure}

\subsection{Temperature dependence of $S$ and $PF$}
By substituting the calculated $L_{11}$ and $L_{12}$ values into Eqs.~(\ref{eq:S}) and (\ref{eq:PF}), we can obtain the $S$ and $PF$ values.
Figures~\ref{fig:pf}(a) and \ref{fig:pf}(b) show the $T$ dependences of the $PF$ and $S$ of an FeSe thin film in $\mathcal E_{\perp}$, respectively.
The solid circles are experimental data reported by Shimizu {\it et al.}, and the solid and dashed curves represent the theoretical curves obtained on the basis of the two-band-with-finite-splitting model and the degenerate-two-band model, respectively. 
In Fig.~\ref{fig:pf}(a) we can see that the solid curve shows excellent agreement with the experimental $PF$ data, whereas the dashed curve cannot reproduce the experimental data.
Similarly, the solid curve in Fig.~\ref{fig:pf}(b) reproduces the experimentally obtained high Seebeck coefficient with $S\sim 350~\mu$V/K, whereas the dashed curve deviates substantially from the experimental data.
The band splitting $\Delta$ originating from the $\mathcal E_{\perp}$ is critical for reproducing the large $S$ of an FeSe thin film.
Thus, using the two-band model, we succeeded in explaining the high $PF$ and $S$ of the FeSe thin film mentioned in the introduction.
Note that the shoulder structure at approximately $200$~K in the experimental $S$ data, as shown in Fig.~\ref{fig:pf}(b), is not explained by the present model, which might indicate a change in the electronic states near this temperature.

\section{Several Possible Effects on the TE Response of FeSe Thin Films~\label{sec:4}}
\subsection{Effects of superconducting fluctuations}
As introduced in \S~1, the present FeSe thin film is superconducting below $T_{\rm c}\sim50$~K.
It is interesting to identify possible effects of superconducting fluctuations on $L_{11}$ and $L_{12}$ in the $T$ range where $S$ is large and weakly $T$ dependent.
In studies on dirty superconductors, two distinct processes---the Azlamasov--Larkin process~\cite{rf:Aslamazov01, rf:Aslamazov02} and the Maki--Thompson process~\cite{rf:MK,rf:thompson}---have been proposed theoretically and identified experimentally in thin Al films~\cite{rf:crow}.
The Azlamasov--Larkin process is responsible for the main contributions near $T_{\rm c}$, which is given as follows by the conductivity $\sigma'$ arising from the fluctuation for a thin film with thickness $d$:
\begin{eqnarray}
\sigma'=\frac{e^2}{\pi\hbar d}\frac{\pi}{16}\frac{T_{\rm c}}{T-T_{\rm c}}.
\label{eq:sigma'}
\end{eqnarray}
This result is to be compared with the dominant contributions to conductivity, $\sigma_0$, given by
\begin{eqnarray}
\sigma_0=\frac{e^2}{\pi\hbar d}\frac{\tau E}{\hbar}.
\label{eq:sigma'}
\end{eqnarray}
Here, $\sigma_0$ can be determined from the experimental data~\cite{rf:shimizu} ({\it e.g.}, $\tau E /\hbar=5\times10^{11}T^{-5}$); thus, $\sigma'$ is smaller than $\sigma_0$ by an order of $\hbar/\tau E$ in the $T$ range except around $T_{\rm c}$, which then leads to small contributions to $L_{11}$.
The contribution of $\sigma'$ to $L_{12}$ is also small because $\sigma'$ is weakly dependent on energy in the present case.
This fact has been numerically assessed ({\it e.g.}, the weak $T$ dependence of $S$ with a large value of 350~$\mu$V/K for $50~{\rm K}<T<300~{\rm K}$ is difficult to understand in terms of superconducting fluctuations).

\begin{figure}[t]
  \begin{center}
  \includegraphics[keepaspectratio=true,width=80mm]{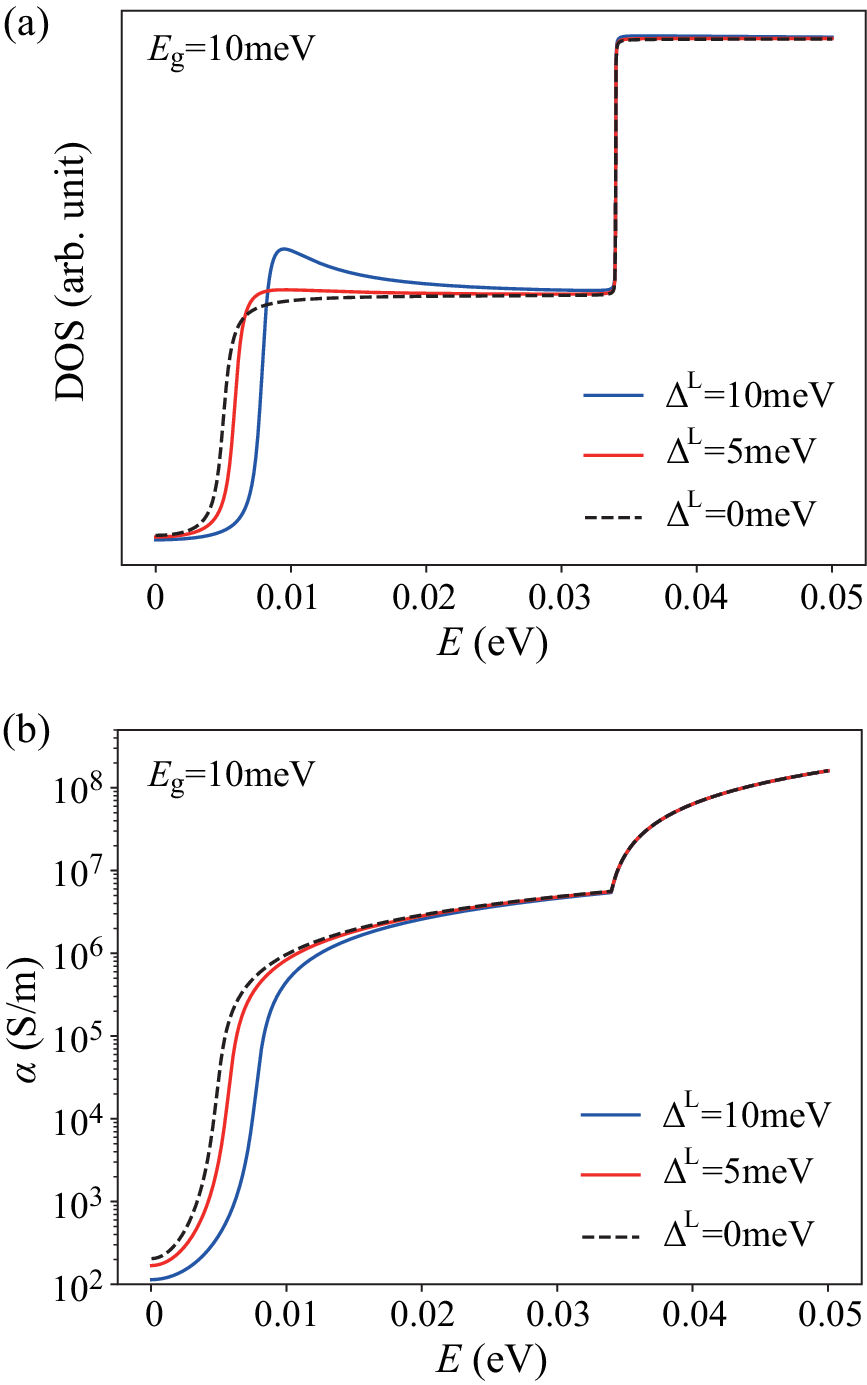}
  \end{center}
\caption{(Color online) Energy dependence of (a) the DOS and (b) the spectral conductivity of an FeSe thin film with a narrow gap $E_{\rm g}=10$~meV for $\Delta^{\rm L}=0$ (dashed curve), $5$~meV (red curve), and 10~meV (blue curve) at $T=100$~K. 
$\Delta_{\rm ex}^{\rm H}$ is zero for all curves.}
\label{fig:rho_ex_L}
\end{figure}

\begin{figure}[t]
  \begin{center}
  \includegraphics[keepaspectratio=true,width=80mm]{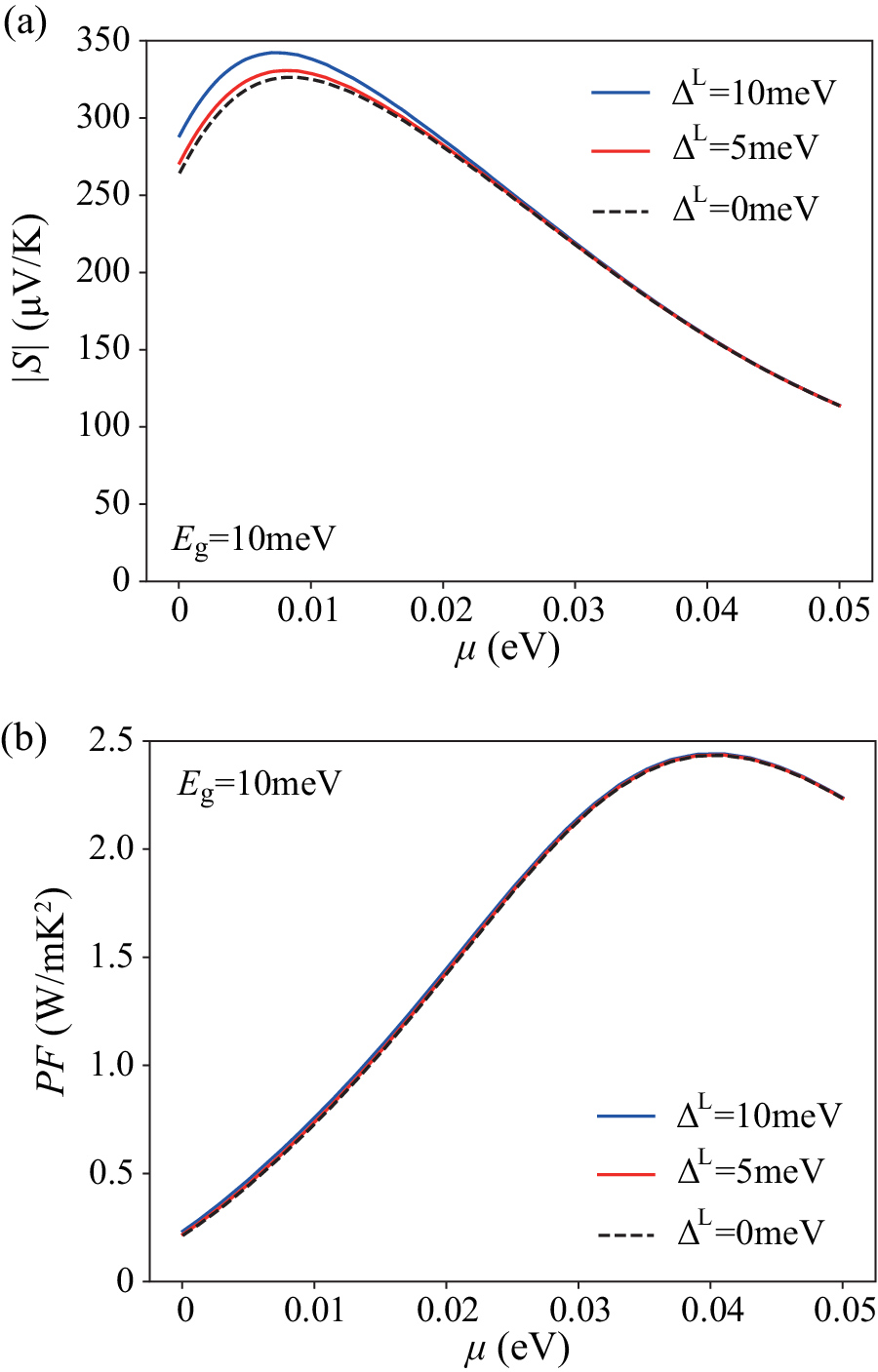}
  \end{center}
\caption{(Color online) The $\mu$ dependence of (a) $S$ and (b) $PF$ of an FeSe thin film with a narrow gap $E_{\rm g}=10$~meV for $\Delta^{\rm L}=0$ (dashed curve), $5$~meV (red curve), and 10~meV (blue curve) at $T=100$~K.
$\Delta^{\rm H}_{\rm ex}$ is zero for all curves.}
\label{fig:pf_ex_L}
\end{figure}

\subsection{Effects of excitonic correlations}
The system of present interest, FeSe, is semimetallic with an equal number of electrons and holes when in bulk form; however, in thin-film form, FeSe has more electrons under an applied gate voltage.
In such a situation, effects of excitonic correlations~\cite{rf:hal}, including orbital ordering~\cite{rf:ono}, are reasonably expected.
To assess these effects, we incorporated the valence band into the two-band model via the following $3\times 3$ Hamiltonian:
\begin{eqnarray}
H(\bm k)=
\begin{pmatrix}
 \dfrac{\hbar^2k^2}{2m}+\dfrac{E_{\rm g}}{2}+\Delta & 0 &\Delta_{\rm ex}^{\rm H}\\
0 &  \dfrac{\hbar^2k^2}{2m}+\dfrac{E_{\rm g}}{2} & \Delta_{\rm ex}^{\rm L} \\
\Delta_{\rm ex}^{\rm H} &  \Delta_{\rm ex}^{\rm L} & -\dfrac{\hbar^2k^2}{2m}-\dfrac{E_{\rm g}}{2} \\
\end{pmatrix}.
\label{eq:ex}
\end{eqnarray}
Here, $E_{\rm g}$ is the energy gap between the lower conduction band and the valence band, $\Delta$ is the energy spacing between the bottoms of the lower and higher conduction bands, and $\Delta_{\rm ex}^{\rm L/H}$ is the excitonic order parameter for the lower/higher conduction band.
$\Delta_{\rm ex}^{\rm L/H}$ is assumed to be expressed as follows by the mean-field solution in mind:
\begin{eqnarray}
\Delta_{\rm ex}^{\rm L/H}=\Delta^{\rm L/H}\tanh\left(\sqrt{\dfrac{T_0-T}{T_0}}\right),
\end{eqnarray}
where $T_0$ is the critical temperature.
In the following calculation, we set $m=2.7m_0$, $E_{\rm g}=10$~meV, $\Delta=29$~meV, and $T_0=200$~K.
We also adopted the constant-$\tau$ approximation to the self-energy, as in Eq.~(\ref{eq:sigma}).
(See the Appendix for the calculation procedure of the TE response in this system; the calculation is based on the KL theory within the constant-$\tau$ approximation.)

We now discuss the two special cases of (i) $\Delta_{\rm ex}^{\rm H}=0$ and (ii) $\Delta_{\rm ex}^{\rm L}=0$ to understand possible effects of excitonic correlations on the lower and higher bands separately.

\paragraph{(i) Case of $\Delta_{\rm ex}^{\rm H}=0$}
Figure~\ref{fig:rho_ex_L} shows the energy dependence of the DOS and the spectral conductivity at $T=T_0/2$=100~K for several values of $\Delta^{\rm L}$.
In this calculation, the relaxation times are chosen to be $\hbar/\tau_{\rm v}=\hbar/\tau_{\rm L}=1$ meV and $\hbar/\tau_{\rm H}=0.02$ meV.
As $\Delta^{\rm L}$ increases, the energy gap between the lower band and the valence band becomes large and the DOS increases near the bottom of the lower band, as shown in Fig.~\ref{fig:rho_ex_L}(a).
As a result, $\alpha(E)$ in Fig.~\ref{fig:rho_ex_L}(b) decreases near the bottom of the lower band.

Figure~\ref{fig:pf_ex_L} represents the $\mu$ dependence of $S$ and $PF$ at $T=100$~K for several values of $\Delta^{\rm L}$.
$S$ increases slightly with increasing $\Delta^{\rm L}$ when $\mu$ locates near the bottom of the lower band.
However, the excitonic effect on $PF$ is negligibly small even near the lower-band bottom.

\begin{figure}[t]
  \begin{center}
  \includegraphics[keepaspectratio=true,width=80mm]{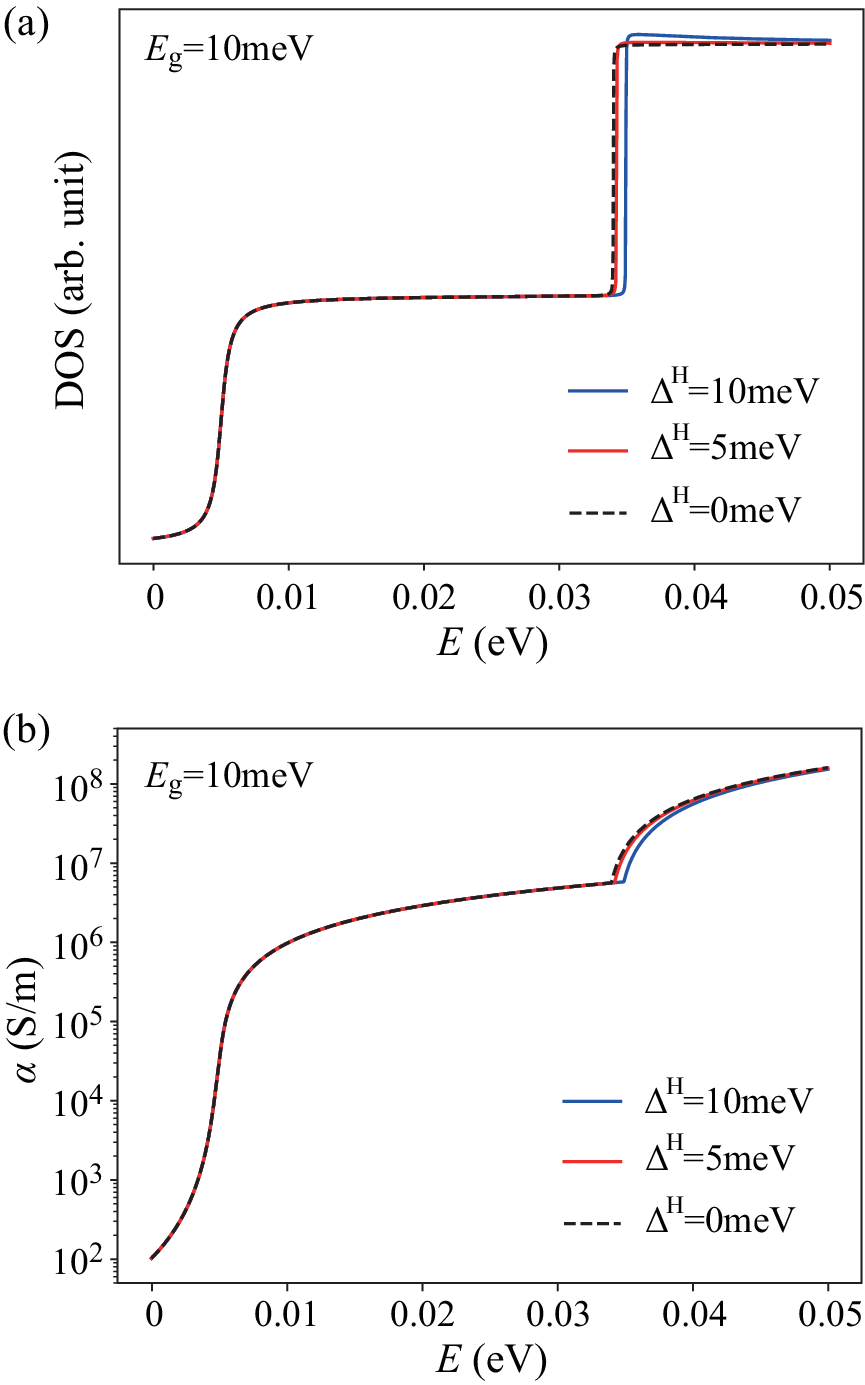}
  \end{center}
\caption{(Color online) Energy dependence of (a) the DOS and (b) the spectral conductivity of an FeSe thin film with a narrow gap ($E_{\rm g}=10$~meV) for $\Delta^{\rm H}=0$ (dashed curve), $5$~meV (red curve), and 10~meV (blue curve) at $T=100$~K.
$\Delta^{\rm L}_{\rm ex}$ is zero for all the curves.}
\label{fig:rho_ex_H}
\end{figure}

\paragraph{(ii) Case of $\Delta_{\rm ex}^{\rm L}=0$}
Figure~\ref{fig:rho_ex_H} shows the DOS and $\alpha(E)$ at $T=100$ K for several values of $\Delta^{\rm H}$.
In this calculation, the relaxation times are set as $\hbar/\tau_{\rm v}=\hbar/\tau_{\rm H}=0.02$~meV and $\hbar/\tau_{\rm L}=1$~meV.
Figure~\ref{fig:rho_ex_H}(a) shows that excitonic effects appear near the bottom of the higher band but are weaker than in the case of $\Delta_{\rm ex}^{\rm H}=0$.
Similarly, excitonic effects on $\alpha(E)$ are also small, as shown in Fig.~\ref{fig:rho_ex_H}(b).
Because of the small change of $\alpha(E)$, the values of $S$ and $PF$ are weakly affected by excitonic interaction.

\section{Summary}
We introduced a simple theoretical model, the two-band model, to attain both a large $S$ and high $L_{11}$ and thereby obtain a high $PF$ and then applied it to an FeSe thin film under $\mathcal E_{\perp}$.
We succeeded in understanding the recent experimental report by Shimizu {\it et al.}~\cite{rf:shimizu}, who found that an FeSe thin film in the presence of $\mathcal E_{\perp}$ exhibits a high $PF$ originating from a large $S$ and high $L_{11}$.
In the case of an FeSe thin film, $\mu$ locates near the bottom of the lower conduction band, resulting in $S\sim 350~\mu$V/K and $PF\sim 500$~mW/(m$\cdot$K$^2$) at $T\sim 100$~K.
We also examined the effects of superconducting fluctuations and excitonic correlations, and we conclude that their effects are not critical for the high $PF$ of an FeSe thin film.

Similar candidates with a high $PF$ resulting from the two bands are nanowires and nanotubes with nanosized diameters because their energy bands have sub-band structures as a result of quantum confinement effects along the diameter direction.
Notably, the sub-band gap in these candidates is independent of $\mathcal E_{\perp}$ and only $\mu$ can be controlled by tuning $\mathcal E_{\perp}$, in contrast to FeSe thin films. 

\section*{Acknowledgements}
We thank Yoshihiro Iwasa for the fruitful discussions on experiments on FeSe thin films.
This work was partly supported by JSPS KAKENHI (Grant No. 22K18954).

\appendix
\section{Excitonic effects on the DOS and the spectral conductivity}
In this Appendix, we represent analytical expressions of the DOS and the spectral conductivity for the two-band model incorporating the effects of excitonic correction, which is described by the Hamiltonian in Eq.~(\ref{eq:ex}).
In the following discussion, we focus on the case of $\Delta_{\rm ex}^{\rm H}=0$, as discussed in the main text.

We adopt the constant-$\tau$ approximation for the self-energy as
\begin{eqnarray}
\Sigma^{\rm R/A}=
\begin{pmatrix}
\mp i\hbar/2\tau_{\rm H} & 0 &0\\
0& \mp i\hbar/2\tau_{\rm L} &0\\
0&0&\mp i\hbar/2\tau_{\rm v} \\
\end{pmatrix},
\label{eq:ex_sigma}
\end{eqnarray}
where we set $\tau_{\rm L} = \tau_{\rm v}$.
The DOS is described as
\begin{eqnarray}
\rho(E)=-g_{\rm s}g_{\rm o}
\frac{1}{\pi\Omega}\Tr\left(\sum_{\bm k}{\rm Im} G^{\rm R}(E,{\bm k})\right)
\end{eqnarray}
in terms of the retarded Green's function $G^{\rm R}(E,{\bm k})=\left[EI-H({\bm k})-\Sigma^{\rm R}\right]^{-1}$ and can be analytically calculated as
\begin{eqnarray}
\rho(E)=\rho_{\rm H}(E)+\rho_{\rm L-v}(E)
\label{eq:self-energy_A1}
\end{eqnarray}
with
\begin{eqnarray}
\rho_{\rm H}(E)
=g_{\rm s}g_{\rm o}\frac{m}
{2\pi^2\hbar^2}\left(\pi-\Theta_{\rm H}\right)
\end{eqnarray}
and
\begin{eqnarray}
\rho_{\rm L-v}(E)
=g_{\rm s}g_{\rm o}
\frac{m}{2\pi^2\hbar^2}{\rm Im}\left[
\frac{|E|+i\frac{\hbar}{2\tau_{\rm L}}}{z^{\rm R}}
\left\{
\ln \frac{\left|
\frac{E_{\rm g}}{2}+z^{\rm R}\right|}{\left|\frac{E_{\rm g}}{2}-z^{\rm R}\right|}\right.\right.\notag\\
\left. \left.
+i(\pi-\Theta_{\rm L-v}^{(-)}+\Theta_{\rm L-v}^{(+)})
\right\}
\right].
\end{eqnarray}
Here, $\rho_{\rm H}(E)$ is the DOS of the higher band without excitonic correlations and $\rho_{\rm L-v}(E)$ is the DOS of the lower conduction band and the valence band including excitonic correlations. 
$z^{\rm R/A}$ are given by
\begin{eqnarray}
z^{\rm R/A}=\sqrt{\left(|E|\pm i\hbar/2\tau_{\rm L}\right)^2-\left(\Delta^{\rm L}_{\rm ex}\right)^2}
\label{eq:self-energy_A2}
\end{eqnarray}
with $ {\rm Im}\ z^{\rm R}\geq0$ and  ${\rm Im}\ z^{\rm A}\leq0$.
$\Theta_{\rm H}$ and $\Theta_{\rm L-v}^{(\pm)}$ are defined as
\begin{eqnarray}
\Theta_{\rm H}=\arctan\frac{\hbar}{2\tau_{\rm H}\left(E-\frac{E_{\rm g}}{2}-\Delta\right)}\ \ \ (0\leq\Theta_{\rm H}<\pi),\\
\Theta_{\rm L-v}^{(\pm)}=\arctan\frac{{\rm Im}\ z^{\rm R}}{{\rm Re}\left(z^{\rm R}\pm\frac{E_{\rm g}}{2}
\right)}\ \ \ (0\leq\Theta_{\rm L-v}^{(\pm)}<\pi).
\label{eq:Theta}
\end{eqnarray}

The spectral conductivity $\alpha(E)$ is expressed as 
\begin{eqnarray}
\alpha(E)&=&g_{\rm s}g_{\rm o}\frac{\hbar e^2}{2\pi V}\sum_{\bm k}{\rm Re}\Tr\left[v_xG^{\rm A}(E,{\bm k})v_xG^{\rm R}(E,{\bm k})\right.\nonumber\\
&&\left.-v_xG^{\rm R}(E,{\bm k})v_xG^{\rm R}(E,{\bm k})\right]
\label{eq:alpha_E}
\end{eqnarray}
in terms of the retarded/advanced Green's functions $G^{\rm R/A}(E,{\bm k})$ and the velocity matrix $v_x=\hbar^{-1}\partial H({\bm k})/\partial k_x$ in the $x$-direction along the temperature gradient of the system.
Within the constant-$\tau$ approximation, Eq.~(\ref{eq:alpha_E}) can be calculated as
\begin{eqnarray}
\begin{split}
\alpha(E)=\alpha_{\rm H}(E)+\alpha_{\rm L-v}(E)
\label{eq:alpha_ex}
\end{split}
\end{eqnarray}
with the contribution from the higher conduction band,
\begin{eqnarray}
\alpha_{\rm H}(E)
=g_{\rm s}g_{\rm o}\frac{e^2}{4\pi^2\hbar d}
\left\{
\frac{2\tau_{\rm H}}{\hbar}\left(
E-\frac{E_{\rm g}}{2}-\Delta
\right)
(\pi-\Theta_{\rm H})+1\right\}
\label{eq:alpha_H}
\end{eqnarray}
and the contribution from both the lower conduction band and the valence band,
\begin{eqnarray}
\begin{split}
&\alpha_{\rm L-v}(E)
=g_{\rm s}g_{\rm o}\frac{e^2}{4\pi^2\hbar d}
\left[
\frac{2\tau_{\rm L}}{\hbar |E|}
\left\{
\left(E^2-\left(\Delta^{\rm L}_{\rm ex}\right)^2\right)
\left(\pi-\Theta_{\rm L-v}^{(-)}-\Theta_{\rm L-v}^{(+)}\right)
\right.
\right.\\
&\left.
\left.
-\frac{E_{\rm g}}{2}{\rm Im}
\left[
\frac{E^2-\left(\Delta^{\rm L}_{\rm ex}\right)^2+i\frac{\hbar}{2\tau_{\rm L}}|E|}{z^{\rm R}}
\left\{
\ln \frac{\left|
\frac{E_{\rm g}}{2}+z^{\rm R}\right|}{\left|\frac{E_{\rm g}}{2}-z^{\rm R}\right|}
+i\left(\pi-\Theta_{\rm L-v}^{(-)}+\Theta_{\rm L-v}^{(+)}\right)
\right\}
\right]
\right\}
+2
\right].
\label{eq:alpha_L}
\end{split}
\end{eqnarray}
The DOS and the spectral conductivity for the case of $\Delta_{\rm ex}^{\rm L}=0$ are obtained similarly.

\end{document}